\documentclass{emulateapj}
\newcommand{\kel}{\mbox{ K}}
\newcommand{\mkel}{\mbox{ mK}}

\newcommand{\Mpcinv}{\mbox{ Mpc$^{-1}$}}

\newcommand{\hunits}{\mbox{ km s$^{-1}$ Mpc$^{-1}$}}
\newcommand{\bq}{\begin{equation}}
\newcommand{\eq}{\end{equation}}
\newcommand{\bqa}{\begin{eqnarray}}
\newcommand{\eqa}{\end{eqnarray}}

\newcommand{\xh}{x_{\rm HI}}
\newcommand{\bxion}{\bar{x}_i}
\newcommand{\mmin}{m_{\rm min}}
\newcommand{\mmax}{m_{\rm max}}
\newcommand{\dTb}{\delta T_b}
\newcommand{\bdTb}{\bar{\delta T}_b}
\newcommand{\Junits}{\mbox{ cm$^{-2}$ s$^{-1}$ Hz$^{-1}$ sr$^{-1}$}}
\newcommand{\Pmh}{P_{\rm mh}}
\newcommand{\fmh}{f_{\rm mh}}
\newcommand{\deriv}{d}
\newcommand{\lya}{Ly$\alpha$ }

\begin{document}

\title{Redshifted 21 Centimeter Emission from Minihalos Before Reionization}

\author{Steven R.  Furlanetto\altaffilmark{1} \& S.~Peng Oh\altaffilmark{2}}

\altaffiltext{1} {Yale Center for Astronomy and Astrophysics, Yale University, 260 Whitney Avenue, New Haven, CT 06520-8121; steven.furlanetto@yale.edu}

\altaffiltext{2}{Department of Physics, University of California, Santa Barbara, CA 93106;  peng@physics.ucsb.edu}

\begin{abstract}
Before reionization, the intergalactic medium (IGM) may have been sufficiently cold for low-mass ``minihalos" to condense out of the gas and subsequently affect reionization.  Previous work has shown that minihalos generate reasonably large 21 cm fluctuations.  Here we consider this signal in its proper cosmological context and show that isolating minihalos from the rest of the IGM is extremely difficult.  Using the well-known halo model, we compute the power spectrum of 21 cm fluctuations from minihalos and show that the signal decreases rapidly as feedback increases the Jeans mass.  We then show that even a small Ly$\alpha$ background increases the 21 cm fluctuations of the diffuse IGM well beyond those of the minihalos; because the mass fraction in the IGM is much larger, minihalos will lie buried within the IGM signal. The distinctive signatures of non-linear bias and minihalo structure emerge only at much smaller scales, well beyond the resolution of any upcoming instruments. Using simple, but representative, reionization histories, we then show that the required Ly$\alpha$ background level is most likely achieved at $z \ga 15$, while minihalos are still rare, so that they are almost always degenerate with the diffuse IGM. 
\end{abstract}
  

\section{Introduction}
\label{intro}

Cosmologists have long recognized that the ubiquity of hydrogen in the intergalactic medium (IGM) makes its line transitions particularly effective diagnostics of the diffuse material that separates galaxies.  One especially useful probe is the so-called Ly$\alpha$ forest \citep{gunn65}, which has proved exceedingly useful for studying the $z \la 6$ Universe.  Unfortunately, this transition is so strong that the forest saturates at relatively small neutral fractions (e.g., \citealt{fan06}).  Thus the much weaker 21 cm hyperfine transition is the probe of choice for studying the predominantly neutral gas in the high-redshift IGM, especially during and before reionization \citep{field59-obs, sunyaev75, hogan79, scott90, madau97}.  

Because it is a spectral line, 21 cm observations offer the possibility of mapping the three dimensional distribution of gas from $z \sim 200$, when the gas temperature first decoupled from the CMB, to $z \sim 6$, a period spanning the ``Dark Ages," the onset of structure formation, the birth of the first galaxies, and reionization itself.  One particularly interesting application is to study the first collapsed objects.  The IGM is visible only so long as the 21 cm spin temperature $T_S$ differs from the CMB temperature $T_\gamma$ (because the latter is used as a backlight).  At $z \la 50$ (and before luminous sources appear), the vast majority of gas is invisible because collisions are not rapid enough to drive $T_S \rightarrow T_K$, the kinetic temperature of the gas (which is colder than $T_\gamma$ because of adiabatic expansion).  Only overdense regions remain visible.  \citet{iliev02, iliev03} pointed out that ``minihalos," which are dark matter halos larger than the Jeans mass but too small to cool and form stars, constitute a particularly interesting class of 21 cm emitters.  These objects are a qualitatively new feature of the pre-reionization IGM that can slow reionization because of their rapid recombination rates \citep{haiman01, iliev05, ciardi05-mh}.  Although they are much too small to be imaged individually in realistic experiments, they imprint large-scale fluctuations on the 21 cm background that could, in principle, be used to constrain cosmological parameters \citep{iliev02, iliev03}.

The purpose of this paper is to consider the minihalo signal in more detail and to embed it in its proper cosmological context.  Because minihalos cannot be resolved, they are obviously subject to confusion with other sources in the beam.  \citet{furl04-sh} pointed out one possibility:  collisions can be sufficiently rapid in moderately overdense IGM gas to render visible sheets and filaments in the cosmic web. This has since been confirmed in numerical simulations \citep{kuhlen06-21cm, shapiro05}, where sheets and filaments contribute a significant fraction of the signal at $z \ga 15$ and confuse attempts to identify the minihalos.  But here we will focus on the low-density, diffuse IGM.  The Wouthuysen-Field mechanism, in which absorption and re-emission of Ly$\alpha$ photons mixes the spin states \citep{wouthuysen52, field58, hirata05} can drive $T_S \rightarrow T_K$ everywhere and hence render the entire IGM visible.  \citet{oh03} argued that, if Wouthuysen-Field coupling is strong, the minihalos (which contain only a small fraction of the mass)  will be buried in the rest of the signal.  

A related question is how efficiently minihalos can form once feedback from the first luminous sources is taken into account.  In particular, X-ray heating increases the cosmological Jeans mass.  The resulting ``entropy floor" prevents gas accretion onto minihalos \citep[hereafter OH03]{oh03-entropy},  effectively suppressing their abundance.  But the importance of this mechanism is unclear, because X-rays can also stimulate cooling, and entropy injection will be less efficient in minihalos that have already begun to collapse.  One of the key questions of reionization is the efficiency of feedback, and 21 cm observations of minihalos could potentially shed light on it.  Here we will develop a new method to calculate the 21 cm fluctuations of minihalos (based on the halo model) and use it to predict the observable implications of feedback.

The rest of this paper is organized as follows.  We begin in \S \ref{mh} and \S \ref{21cm} with brief discussions of minihalos and 21 cm physics.  In \S \ref{flucs}, we compute the fluctuation amplitude expected from minihalos as a function of the IGM entropy.  We then relate that signal to fluctuations from the diffuse IGM in \S \ref{igm}, including Ly$\alpha$ coupling.   Next we consider minihalos in the context of some representative reionization histories in \S \ref{hist}.  Finally, we conclude in \S \ref{disc}.

In our numerical calculations, we assume a cosmology with $\Omega_m=0.26$, $\Omega_\Lambda=0.74$, $\Omega_b=0.044$, $H=100 h \hunits$ (with $h=0.74$), $n=0.95$, and $\sigma_8=0.8$, consistent with the most recent measurements \citep{spergel06}.\footnote{Note that we have increased $\sigma_8$ above the best fit \emph{WMAP} value in order to improve agreement with weak lensing measurements.}  Unless otherwise specified, we use comoving units for all distances.

\section{Minihalos and Feedback}
\label{mh}

While collisionless cold dark matter can collapse into gravitationally bound objects of arbitrary masses, thermal pressure prevents baryons from accreting onto a halo unless its mass exceeds the time-averaged Jeans mass, or the ``filter" mass \citep{gnedin98}.  We take this, evaluated in an adiabatically cooling universe, to be the minimum allowable mass $\mmin$ for a minihalo.  We set the maximum mass $\mmax$ to correspond to a virial temperature $T_{\rm vir}=10^4 \kel$ (defined as in \citealt{barkana01}), the point at which atomic cooling becomes efficient.  

Of course, if the IGM temperature exceeds its nominal value from adiabatic cooling, the Jeans mass increases and minihalo formation is suppressed.  OH03 showed that the IGM ``entropy" $K_{\rm IGM} \equiv T_K/n^{2/3}$ quantifies the suppression; here $T_K$ is the kinetic temperature of the IGM.  Clearly $K$ is conserved so long as the gas evolves adiabatically, which occurs both during the overall expansion of the universe and during the initial stages of halo collapse.  Subsequent shock heating can further raise the gas entropy, but so long as radiative cooling is unimportant, $K_{\rm IGM}$ provides a well-defined entropy floor.  We can similarly define $K_{\rm mh}(r) = T_{\rm vir}/n(r)^{2/3}$ to be the entropy profile acquired by a minihalo during gravitational collapse. We assume that the gas density profile $n(r)$ traces the universal dark matter profile of \citet{navarro97}.  Numerical simulations (e.g., \citealt{frenk99}) show this to be a good assumption except near the halo center, where in any case the effects of a finite entropy floor predominate. 
 
With this entropy profile, it is possible to calculate detailed equilibrium density and temperature profiles. Indeed, by constructing these profiles as a function of $K_{\rm IGM}$, OH03 showed that the structure and total accreted gas fraction $f_g$ (relative to the cosmic mean $\Omega_b/\Omega_m$) are simply functions of $\hat{K} \equiv K_{\rm IGM}/K_{\rm mh}(r_{\rm vir})$ (see their Fig.~6).  If $\hat{K} \ll 1$, the entropy generated by the gravitational accretion shock vastly exceeds the IGM entropy and $f_g \rightarrow 1$.  If, on the other hand, $\hat{K} \ga 1$, the thermal pressure of the IGM exceeds the minihalo's gravitational attraction, so $f_g \rightarrow 0$.  We compute $f_g$ for each minihalo by interpolating the results of OH03.  Note that these gas fractions match the fitting formula of \citet{gnedin00-mfil} reasonably well; the widely-used filter mass is simply a crude estimate of the point at which $\hat{K}=1$ in an adiabatically cooling universe.

This prescription for calculating density profiles (and hence $f_{g}$) assumes that $K_{\rm IGM}$ is constant throughout the accretion history of the halo; in reality, $K_{\rm IGM}$ depends on the thermal history of the minihalo's progenitors, but we shall ignore such complications here. Furthermore (as in this paper), radiative cooling by ${\rm H_{2}}$ molecules was assumed to be negligible. \citet{rozas06} obtain comparable results with a similar model. Note that if the soft-UV background flux is small, the effects of radiative cooling, which erases the entropy floor established by an X-ray background, cannot be ignored \citep{kuhlen05-sim}. Generally, such considerations only apply to extremely hard spectra. 

\section{The 21 cm Transition}
\label{21cm}

We review the relevant characteristics of the 21 cm transition here; we refer the interested reader to \citet{furl06-review} for a more comprehensive discussion.  The 21 cm brightness temperature (relative to the CMB) of a patch of the IGM is
\begin{eqnarray}
\dTb & = & 27 \xh (1 + \delta) \, \left( \frac{\Omega_b h^2}{0.023} \right) \left( \frac{0.15}{\Omega_m h^2} \, \frac{1+z}{10} \right)^{1/2} \nonumber \\
& & \times \left( \frac{T_S - T_\gamma}{T_S} \right) \mkel,
\label{eq:dtb}
\end{eqnarray}
where $\xh$ is the neutral fraction, $\delta$ is the fractional overdensity, $T_\gamma$ is the CMB temperature at redshift $z$, and $T_S$ is the spin temperature.  Note that the patch will appear in absorption if $T_S < T_\gamma$ and emission otherwise.  Here we have assumed that the patch expands uniformly with the Hubble flow; radial peculiar velocities also affect $\dTb$ by changing the mapping from distance to frequency \citep{bharadwaj04-vel,barkana05-vel}.  However, these anisotropic fluctuations do not affect our argument, so we will ignore them for simplicity.\footnote{Note that the velocity field around an individual minihalo is far from the Hubble flow, so this is not a good assumption if the instrument is able to resolve the halos.  However, such high-resolution experiments will be impractical for the foreseeable future.}

The spin temperature $T_S$ is determined by the competition between three processes:  scattering of CMB photons, collisions, and scattering of Ly$\alpha$ photons \citep{wouthuysen52,field58}.  In equilibrium,
\begin{equation}
T_S^{-1} = \frac{T_\gamma^{-1} + \tilde{x}_\alpha \tilde{T}_c^{-1} + x_c T_K^{-1}}{1 + \tilde{x}_\alpha + x_c}.
\label{eq:tsdefn}
\end{equation}
Here $x_c \propto n$ is the collisional coupling coefficient \citep{zygelman05}, $T_{K}$ is the gas kinetic temperature, and $\tilde{T}_{c}$ is the effective color temperature of the radiation field.  Normally $\tilde{T}_c \approx T_K$ \citep{field59-ts, hirata05}.  The middle term describes the Wouthuysen-Field effect, in which absorption and re-emission of Ly$\alpha$ photons mixes the hyperfine states.  The coupling coefficient is \citep{chen04}
\bq
\tilde{x}_\alpha = 1.81 \times 10^{11} (1+z)^{-1} \tilde{S}_\alpha J_\alpha,
\label{eq:xalpha}
\eq
where $\tilde{S}_\alpha$ is a factor of order unity describing the detailed atomic physics of the scattering process and $J_\alpha$ is the background flux at the Ly$\alpha$ frequency in units of $\Junits$; the Wouthuysen-Field effect becomes efficient when there is about one Ly$\alpha$ photon per ten baryons.  We use the numerical fits of \citet{hirata05} for $\tilde{S}_\alpha$ and $\tilde{T}_c$.

\section{Fluctuations from Minihalos}
\label{flucs}

We will begin by calculating the power spectrum of the 21 cm brightness temperature assuming that \emph{only} minihalos are allowed to emit.  As we will see below, this requires an unlikely set of circumstances, but it provides some intuition about the minihalo signal and the effects of feedback upon it.  We will use the halo model (e.g., \citealt{cooray02}), which constructs the power spectrum of density fluctuations by dividing the universe into dark matter halos of all sizes.  The power spectrum has two components:  a ``one-halo" term that describes correlations between particles within the same virialized halo and a ``two-halo" term that describes correlations between particles in separate halos.  For our purposes, the advantage of this approach is that the contribution from minihalos is obvious:  we simply include only those particles that lie within minihalos.  Thus $\Pmh(k) = \Pmh^{1h}(k) + \Pmh^{2h}(k)$, where
\bqa
\Pmh^{1h}(k) & = & \bdTb^2 \int_{\mmin}^{\mmax} dm \, \left( \frac{m}{\bar{\rho}} \right)^2 \left[ n_h(m)  \, f_g^2(m) \right. \nonumber \\
& & \times \left. {\mathcal T}_{\rm mh}^2(m) \, | u(k|m) |^2 \right], \label{eq:ps1h}
\\
\Pmh^{2h}(k) & = & \bdTb^2 P_{\rm lin}(k) I_{\rm mh}^2(k), \\ 
I_{\rm mh}(k) & = & \int_{\mmin}^{\mmax} dm \, \left( \frac{m}{\bar{\rho}} \right) \left[ n_h(m) \, b(m) \, f_g(m) \right. \nonumber \\
& & \times \left. {\mathcal T}_{\rm mh}(m) \, u(k|m) \right].
\label{eq:ps2h}
\eqa
Here $\bdTb$ is evaluated via equation (\ref{eq:dtb}) with $T_S \gg T_\gamma$ and $\delta=0$, $\bar{\rho}$ is the mean density of the universe, $n_h(m)$ is the halo number density \citep{press74},\footnote{Using the \citet{sheth99} mass function does not affect our conclusions.} $u(k|m)$ is the Fourier transform of the normalized halo density profile, $b(m)$ is the linear bias \citep{mo96}, and ${\mathcal T}$ is a temperature factor.  Although the baryonic distribution differs slightly from the dark matter distribution because of its finite pressure, it suffices to let $u(k|m)$ equal the \citet{navarro97} profile for our purposes.  As we shall see, this only affects the power spectrum on extremely small scales.  

The remaining factor is ${\mathcal T}_{\rm mh}$, which  is the temperature factor $(T_S-T_\gamma)/T_S$ averaged over each minihalo.  \citet{iliev02,iliev03} computed the detailed brightness temperature profiles of minihalos (see also \citealt{furl02-21cm}).  They showed that, although $T_S$ does vary across the minihalo, most gas still has $T_S \gg T_\gamma$ because it is so dense everywhere.  In this case, ${\mathcal T}_{\rm mh} \approx 1$.  For simplicity, we will therefore evaluate ${\mathcal T}_{\rm mh}$ at the mean overdensity of the minihalo ($f_g \Delta_{\rm vir}$, where the latter factor is the mean overdensity of a virialized object; \citealt{bryan98}).  Note that this prescription includes (approximately) the effect of the entropy floor on the 21 cm signal; OH03 make more detailed estimates of the spin temperature profiles of individual minihalos by including the full equilibrium radial density profiles.  This approximation does not affect our conclusions.

\begin{figure}
\plotone{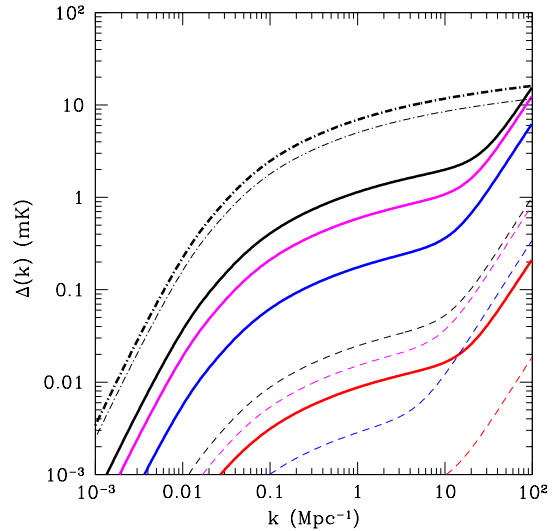}
\caption{Expected rms brightness temperature fluctuations from minihalos.  Thick solid and thin dashed curves are for $z=10$ and $20$, respectively.  From top to bottom within each set, the curves assume $T_K=T_{\rm ad},\, 20,\,100$, and $1000 \kel$ (we set $J_\alpha=0$ throughout).  The dot-dashed curves show the fluctuations from linear theory, assuming that the entire IGM has $T_S \gg T_\gamma$.}
\label{fig:mhbasic}
\end{figure}

Figure~\ref{fig:mhbasic} shows the resulting rms temperature fluctuations as a function of scale, defined via $\Delta^2(k) \equiv k^3 \Pmh(k)/2 \pi^2$.  The thick solid and thin dashed curves show results for $z=10$ and $z=20$, respectively.  We assume $J_\alpha=0$ here, so that Ly$\alpha$ coupling is insignificant.  For reference, the upper thick and lower thin dot-dashed curves show the (linear) amplitude if the entire IGM had $T_S \gg T_\gamma$.  

First consider the uppermost minihalo curve in each set.  These assume that $T_K=T_{\rm ad}$, the temperature appropriate for adiabatic cooling ($T_{\rm ad}=2.58 [(1+z)/10]^{2} \kel$ according to RECFAST; \citealt{seager99, seager00}).  Our choice for $m_{\rm min}$ already includes the effects of the entropy floor for this temperature history, so $f_g \approx 1$ in all minihalos.  The signal has two components.  At scales $k \ga 20 \Mpcinv$, $\Delta$ increases rapidly.  This comes from $P_{\rm mh}^{1h}$, and it corresponds to scales on which the internal structure of minihalos is resolved.  In reality, finite gas pressure will make this part shallower, but such scales are well beyond the capabilities of any currently planned experiments (e.g., \citealt{bowman05,mcquinn05-param}).  Note that our version of the halo model includes only linear bias.  \citet{iliev03} have shown that nonlinear bias modifies the power spectrum at $k \ga 6 \Mpcinv$ -- still somewhat beyond the capabilities of any of these observatories.  On larger scales, the fluctuations trace the linear power spectrum, with the input parameters affecting only the amplitude.  This is the regime where $P_{\rm mh}^{2h}$ dominates.  On such scales, $u(k|m) \approx 1$, so $I_{\rm mh} \rightarrow \bar{b} \fmh$, where $\bar{b}$ is the average (mass-weighted) bias of the minihalos and $\fmh$ is the fraction of the IGM mass incorporated in minihalos.  We find that $(\fmh, \, \bar{b},\, I_{\rm mh}) \approx (0.067,\,2.8,\,0.19)$ and $(6.7 \times 10^{-4},\,8.0,\,0.0054)$ at $z=10$ and $20$, respectively.  Note that, because they are far out on the nonlinear mass tail, the minihalo bias evolves by a large factor over these redshifts.  Nevertheless, minihalo fluctuations are considerably smaller than those from a uniformly hot IGM.  We predict rms fluctuations $\la 2 \mkel$ (or $0.01 \mkel$) at $z=10$ (or 20) on scales accessible to observations.  The signal increases rapidly with cosmic time until $z \sim 8$, when minihalos become nearly linear fluctuations; at significantly higher redshifts they contain such a small fraction of the baryons that detecting them will be a true challenge

These uppermost curves ignore the possibility of thermal feedback in the IGM.  The other curves show the effect of this ``entropy floor":  they take $T_K=20,\,100,$ and $1000 \kel$, from top to bottom.  At the minimum minihalo mass, these have $\hat{K}=(4.9,\,24,\,244)$ and $(2.3,\,11,\,114)$ for $z=10$ and $z=20$.  Obviously even a modest amount of heating dramatically reduces the minihalo signal, especially at $z=10$.  This is because most of the mass is contained in halos with $m \approx m_{\rm min}$, which are most sensitive to feedback.  Thus the minihalo 21 cm signal is probably only detectable if the IGM is relatively cold at the time of their formation.  Their sensitivity to $T_K$ suggests that constraining the minihalo signal could be a powerful probe of feedback, but it also  compromises attempts to use it to constrain cosmological parameters \citep{iliev02}.

\section{Minihalos and the IGM}
\label{igm}

Of course, short of resolving minihalos, any observed patch of sky must contain a mix of low-density IGM gas and minihalos.  To better mimic the actual observables, we now compute the fluctuations from minihalos together with those of the surrounding diffuse IGM.  In the halo model, this IGM gas is made up of all halos with masses between zero and $\mmin$.  Because such halos are too small to accrete gas, the baryons remain at a constant density, so $u(k|m)=1$.  Additionally, the one-halo term vanishes.  The temperature factor ${\mathcal T}_{\rm IGM}$ can be evaluated at the mean cosmic density,\footnote{Actually, in the presence of an entropy floor, the IGM density is slightly larger than the cosmic mean because of gas that could not accrete onto minihalos; we assume that this gas is distributed uniformly throughout the universe.} given a temperature history and a Ly$\alpha$ background.  The total power spectrum of 21 cm fluctuations is then
\bq
P_{21}(k) = \Pmh^{1h}(k) + \bdTb^2 P_{\rm lin}(k)[ I_{\rm mh}(k) + I_{\rm IGM}(k) ]^2,
\label{eq:netfluc}
\eq
where
\bq
I_{\rm IGM}(k) = {\mathcal T}_{\rm IGM} \int_{0}^{\mmin} dm \, n_h(m) \, \left( \frac{m}{\bar{\rho}} \right) \, b(m).
\label{eq:I_IGM}
\eq
The second part is the usual two-halo term split into the IGM, minihalos, and the correlations between them.  The temperature factors are crucial, because if $T_K < T_\gamma$ the IGM absorbs CMB photons while the minihalos emit.  In this case the two tend to cancel each other out.  One consequence of equation (\ref{eq:netfluc}) is that, unless one can reach scales where $P_{\rm mh}^{1h}$ is large, the IGM and minihalo fluctuations are completely degenerate.  Because both trace $P_{\rm lin}$, any minihalo signal can be mimicked by a suitable choice for ${\mathcal T}_{\rm IGM}$.  Note that we have ignored halos with $m>\mmax$ (which are able to cool rapidly and form stars).  If they contain neutral gas, they would boost the signal slightly. 

\begin{figure}
\plotone{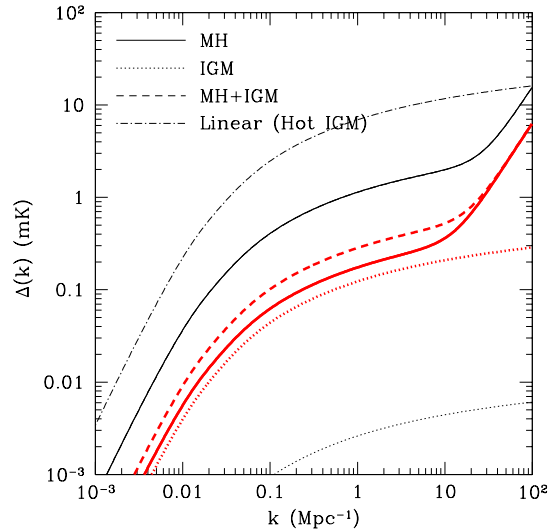}
\caption{Expected rms brightness temperature fluctuations at $z=10$ for $J_\alpha=0$.  The thin and thick curves have $T_K=T_{\rm ad}$ and $T_K=100 \kel$.  Within each set, the solid and dotted lines show the fluctuations from minihalos and the IGM, respectively, while dashed lines show the net result (for $T_K=T_{\rm ad}$, this overlaps the solid curve).  The dot-dashed curve shows the fluctuations according to linear theory, assuming that the entire IGM has $T_S \gg T_\gamma$.   }
\label{fig:temp}
\end{figure}

Of course, our assumption of a constant density isothermal IGM is much too simplistic.  Structure formation organizes the IGM into hot, overdense sheets and filaments that exhibit stronger coupling between the spin and kinetic temperatures.  Using a simple model of spherical collapse, \citet{furl04-sh} argued that when minihalos are still on the nonlinear tail of the mass function (at $z \ga 15$), these shocks will induce 21 cm fluctuations comparable to those of minihalos.  Numerical simulations clearly show that these networks of filamentary emission produce interesting fluctuations that interfere with the minihalo signal, especially at $z \ga 20$ \citep{kuhlen06-21cm,shapiro05}, and some show that $T_S$ fluctuations in the low-density gas can also be significant \citep{gnedin04}.  Our purpose is to show that minihalos will be extremely difficult to separate from fluctuations in the diffuse IGM; by neglecting its spin temperature and density fluctuations, we are therefore taking the most conservative possible stance.  

Figure~\ref{fig:temp} compares the minihalo and IGM 21 cm fluctuations at $z=10$, assuming $J_\alpha=0$.  First consider the thin curves, which assume $T_K=T_{\rm ad}$.  The dot-dashed curve shows the linear density fluctuations if $T_S \gg T_\gamma$.  The solid curve shows the minihalo component, and the dotted curve shows the IGM component.  Minihalos clearly dominate the signal:  the IGM temperature and density are such that $x_c \ll 1$ at $z=10$.  However, if we increase the IGM temperature to $T_K=100 \kel$, as shown in the thick curves (here the dashed line shows the net fluctuations), the situation changes.  First, the minihalo fluctuations decrease by an order of magnitude because the entropy floor suppresses accretion onto them.  At the same time, the IGM fluctuations strengthen because $x_c$ increases with temperature, even though the density remains (nearly) the same.  Thus if $T_K \ga 100 \kel$, the observed fluctuations no longer directly trace the minihalo population, even if $J_\alpha=0$ (provided feedback is as strong as our model predicts).  At higher redshifts, the crossover occurs with even less heat input because the fraction of gas in minihalos is smaller.  Clearly the possibility of IGM heating can significantly reduce our power to interpret weak 21 cm fluctuations as minihalos.

\begin{figure}
\plotone{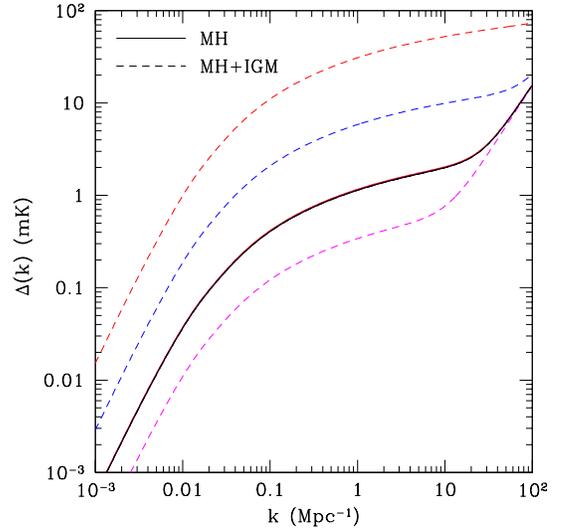}
\caption{Expected rms brightness temperature fluctuations at $z=10$ with $T_K=T_{\rm ad}$.  The solid curve shows the minihalo signal.  The dashed curves show the net 21 cm fluctuations with $J_\alpha= 2 \times (10^{-12},\,10^{-11},\,10^{-10}) \Junits$, from bottom to top.}
\label{fig:ja}
\end{figure}

Thus far we have assumed $J_\alpha=0$ and included only collisional coupling.  Figure~\ref{fig:ja} shows how a Ly$\alpha$ background affects the IGM signal at $z=10$ in a scenario with no X-ray heating.  We emphasize that this maximizes the minihalo component by ignoring the entropy floor.  The solid curve shows the minihalo contribution (which is essentially independent of $J_\alpha$, because collisional coupling is always efficient at virial densities and temperatures).  The dashed curves show the net signal (including the IGM) assuming $J_\alpha=2 \times (10^{-12},\,10^{-11},\,10^{-10}) \Junits$ ($\tilde{x}_\alpha \sim 0.04$--$4$), from bottom to top.  For context, the ionized fraction at a given $\tilde{x}_\alpha$ is \citep{furl06-glob}
\begin{equation}
\bar{x}_{i} \sim 0.005 \left( \frac{1}{1+\bar{n}_{\rm rec}} \, \frac{f_{\rm esc}}{0.1} \, 
\frac{N_{\rm ion}}{N_\alpha} \,
\frac{\tilde{x}_\alpha}{\tilde{S}_\alpha} \, \right) \left( \frac{20}{1+z} \right)^2,
\label{eq:xic}
\end{equation}
where $f_{\rm esc}$ is the escape fraction of ionizing photons, $\bar{n}_{\rm rec}$ is the mean number of recombinations per hydrogen atom, and $N_{\rm ion}/N_\alpha$ is the mean number of ionizing photons per Ly$\alpha$ coupling photon.  This is $\approx 0.45$ for Population II stars and $\approx 7$ for very massive Population III stars.  We thus expect Ly$\alpha$ coupling to turn on well before reionization is complete.  Also note that a soft-UV background of $\tilde{x}_\alpha \sim 0.002$--$0.02$ suffices to suppress ${\rm H_{2}}$ formation and cooling in minihalos \citep{haiman00}, so in principle there can exist a period during which minihalos cannot cool but dominate the 21 cm emission. 

Figure~\ref{fig:ja} shows that, once Ly$\alpha$ coupling begins, the IGM rapidly overwhelms the minihalo signal.  This is because the IGM contains so much more mass than the minihalos and because (unlike emission) absorption does not saturate \citep{oh03}.  We find that, once $\tilde{x}_\alpha \ga 0.02$, the net fluctuations no longer closely trace those of minihalos for $\tilde{x}_{\alpha} \la 0.002$, the minihalos dominate.  As suggested by equation (\ref{eq:xic}) and confirmed in the next section, this occurs well before reionization.  Thus minihalos will only be visible at quite high redshifts.  The Ly$\alpha$ background is the most likely mechanism to render minihalos effectively invisible.  Note as well that the IGM and minihalo signals actually cancel each other out if the IGM is cold.  This is obvious for the bottom curve, but it is also true for the others:  if minihalos did not exist, the rms fluctuations would be larger in both cases.

\section{Minihalos and the Reionization History}
\label{hist}

Thus, minihalos only dominate the observed fluctuations if (1) the IGM remains cold and (2) the Ly$\alpha$ background remains small.  How likely are these conditions to be fulfilled in realistic structure formation models?  In this section, we will use the simple models of \citet{furl06-glob} to consider minihalos in a global context; we refer the reader there for a detailed discussion of the various prescriptions.  In brief, this model assumes that stars dominate the radiation background and sets the total star formation rate via the rate at which gas collapses onto galaxies, $\deriv f_{\rm coll}/\deriv t$, where $f_{\rm coll}$ is the collapse fraction (or the mass in halos with virial temperatures $T_{\rm vir} > 10^4 \kel$).  The ionizing efficiency is $\zeta = A_{\rm He} f_\star f_{\rm esc} N_{\rm ion}$, where $f_\star$ is the star formation efficiency, $f_{\rm esc}$ is the escape fraction of ionizing photons, $N_{\rm ion}$ is the number of ionizing photons produced per baryon incorporated into stars, and $A_{\rm He}$ is a normalization constant accounting for helium in the IGM.  We model recombinations following \citet{miralda00}.

The rate of X-ray heating is also assumed to be proportional to $\deriv f_{\rm coll}/\deriv t$, calibrated to the local relation between star formation rate and X-ray luminosity, measured between $0.2$ and $10$ keV \citep{grimm03,ranalli03,gilfanov04}.  We ignore shock heating.  Finally, we also assume that the soft-UV emissivity (responsible for the Wouthuysen-Field effect) is proportional to $\deriv f_{\rm coll}/\deriv t$ and use the (low-metallicity) Population II and (high-mass) Population III (henceforth PopII/III)spectral fits of \citet{barkana05-ts} to estimate the efficiency with which these photons are produced.  Note that we properly incorporate higher Lyman-series transitions \citep{hirata05, pritchard05}.  

Although there are obviously a number of free parameters in this model, the general features are easy to understand.  The 21 cm background contains three major transitions:  the points when $T_K$ first exceeds $T_\gamma$, when \lya coupling becomes efficient, and when reionization occurs.  Because it affects all the backgrounds equally, $f_\star$ simply shifts everything forward and backward in time.  X-ray heating only affects the kinetic temperature, while $f_{\rm esc}$ only affects the mean ionized fraction $\bxion$.  The stellar initial mass function -- especially the choice between Pop II and very massive Pop III stellar populations -- affects the ratio of \lya photons to ionizing photons and moves the onset of \lya coupling relative to reionization.  Because they are so much hotter, Pop III stars push this transition later in time, compressing the features in the 21 cm background. 

\begin{figure}
\plotone{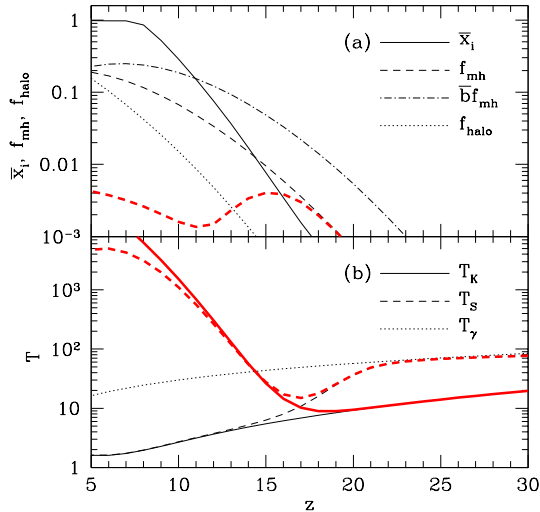}
\caption{Some characteristics of our example Population II history.  \emph{(a)}: The solid line shows $\bxion$, the dotted line shows the collapsed fraction in star-forming halos $f_{\rm coll}$, the dashed lines show the fraction of gas in minihalos $f_{\rm mh}$, and the dot-dashed line shows $\bar{b} f_{\rm mh}$.  \emph{(b)}:  Temperature histories. The thick (or thin) curves include (or ignore) X-ray heating.}
\label{fig:fracs}
\end{figure}

We will use a simple Pop II model.  We take $f_\star=0.1$, $f_{\rm esc}=0.1$, and $N_{\rm ion}=4000$.  The solid curve in Figure~\ref{fig:fracs}\emph{a} shows the resulting ionization history.  We have deliberately chosen $f_\star$ so that reionization occurs at low redshift, thus maximizing the minihalo signal. Note that we have ignored feedback processes during reionization; they will only push the minihalo era further back in time by prolonging reionization.

Figure~\ref{fig:fracs}\emph{b} shows the thermal history.  The thin solid curve in Figure~\ref{fig:fracs}\emph{b} shows $T_K$ if we ignore X-ray heating; the thick solid curve includes it.  Note that, even with our relatively modest injection rate, X-ray heating becomes significant at $z \sim 17$ and rapidly increases the IGM temperature after that point.  The dashed curves show $T_S$.  Interestingly, $T_S \rightarrow T_K$ at $z \sim 20$:  Ly$\alpha$ coupling becomes efficient long before reionization.  This is because only a small fraction of ionizing photons escape their hosts and because Population II stars have a higher specific emissivity in the Lyman-line region than in the Lyman continuum.

Figure~\ref{fig:fracs}\emph{a} also shows the fraction of gas $\fmh$ in minihalos.  Without heating, $\fmh$ increases from $\sim 10^{-3}$ at $z=20$ to $\sim 0.2$ at $z \sim 6$.  But with X-ray heating included, it actually peaks at $z \sim 15$ before plateauing somewhat below $\sim 0.01$.  This is, of course, because the entropy floor prevents gas from accreting onto the dark matter halos.  For reference, the dotted curve shows $f_{\rm coll}$ (note that we neglect the entropy floor for this quantity).

\begin{figure}
\plotone{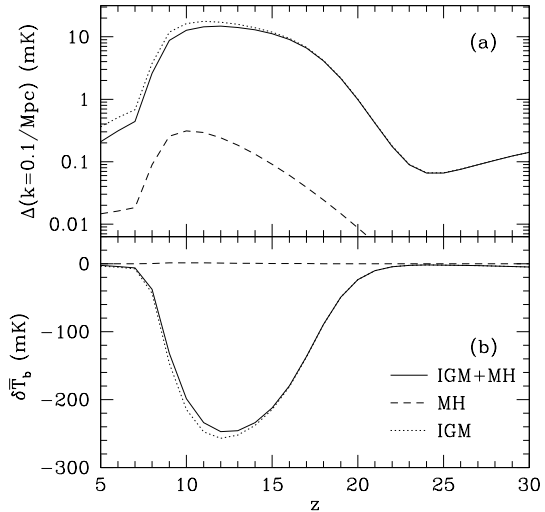}
\caption{Minihalo signal in a cosmological context, for a model with Pop II stars and without X-ray heating.  \emph{(a)}:  Fluctuation amplitude at $k=0.1 \Mpcinv$.  Note that we assume a uniform $\bxion$ here.  \emph{(b)}:  Mean brightness temperature relative to the CMB.  }
\label{fig:pop2_noheat}
\end{figure}

Figure~\ref{fig:pop2_noheat} shows the resulting 21 cm histories if we ignore X-ray heating.  We emphasize that this is \emph{not} realistic and is only meant to build intuition.  Panel \emph{(a)} shows the rms temperature fluctuation at $k=0.1 \Mpcinv$,\footnote{This scale is arbitrary and is chosen only because it sits in the most easily observed window.  The ratio between the curves, which is what is relevant to our argument, is independent of scale so long as the one-halo term can be ignored.} while panel \emph{(b)} shows the mean brightness temperature $\bdTb$.  In each case, the dotted, dashed, and solid curves show the IGM, minihalo, and total signals.  Because the IGM remains cold but Ly$\alpha$ coupling becomes significant at high redshifts, we see a net absorption signal that strengthens toward lower redshifts until reionization. Minihalos on their own contribute $\la 0.3 \mkel$ in emission; however, by removing gas from the strongly absorbing phase, they also reduce $\bdTb$ by a factor $\fmh$.  This becomes significant at $z \la 15$.

Because they are highly clustered, minihalos reduce the fluctuation amplitude by a larger factor ($\bar{b} \fmh$, which can be up to $\sim 30\%$; see the dot-dashed curve in Fig.~\ref{fig:fracs}, which shows this quantity without X-ray heating).  Thus, in principle, minihalos can be detected \emph{indirectly} if both $\bdTb$ and the large-scale rms fluctuation amplitude can be measured.  If only one of the two measurements is available, they are \emph{completely} degenerate with a different ionized fraction, IGM temperature, or cosmology, because the amplitude of each component is quite uncertain.  These separate measurements, while extremely difficult, may in principle be feasible. Radio interferometers are of course insensitive to the temperature zero point, but single-dish measurements may be possible (see, e.g., \citealt{shaver99}). Also, by careful examination of one-point statistics, it may be possible to discern $\bdTb$ to relatively high precision \citep{hansen06}. The expected large-scale density fluctuations are easily computed from linear theory, provided that one can isolate these fluctuations from any other sources (such as ionized bubbles). Alternatively, the fluctuation amplitude can be constrained via redshift-space distortions \citep{barkana05-vel}, although in practice that is rather difficult \citep{mcquinn05-param}. 

\begin{figure}
\plotone{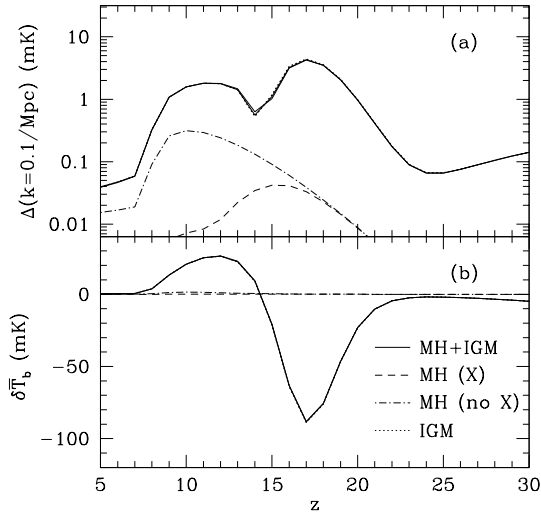}
\caption{Same as Fig.~\ref{fig:pop2_noheat}, except including X-ray heating. The dashed curve shows the signal from minihalos with maximal feedback, while the dot-dashed curve shows the signal neglecting their suppression.  The two solid curves show the net signals in these two cases.  Note that the IGM and total curves overlap nearly everywhere, showing explicitly that minihalos are nearly impossible to isolate without resolving them.  The discontinuity in $\Delta$ corresponds to the crossover from absorption to emission.}
\label{fig:pop2}
\end{figure}

Of course, Figure~\ref{fig:pop2_noheat} is far from realistic because we \emph{must} account for X-ray heating.  Figure~\ref{fig:pop2} shows the results in such a model.  Because the effects of feedback remain controversial, we consider two different prescriptions for it:  one in which feedback maximally suppresses the minihalo abundance (see \S \ref{mh}; dashed lines) and one in which it has no effect on them (dot-dashed lines).  The overall histories are rather different from Fig.~\ref{fig:pop2_noheat}, with an early absorption phase followed by emission once X-ray heating kicks in; at the crossover point, both the mean signal and the fluctuations nearly vanish in our model (note that this is only because we neglect spin temperature fluctuations).  But most importantly, minihalos are now completely swamped by the IGM signal.  \lya coupling turns on at $z \sim 17$, when $f_{\rm mh} \sim 10^{-2}$.  Beyond this point, minihalos become degenerate with the IGM fluctuations.  In principle, combined measurements of $\Delta$ and $\bdTb$ could work when the IGM is in absorption (as in Fig.~\ref{fig:pop2_noheat}), but by $z \sim 15$ the entire IGM is hot and emits relative to the CMB.  Past this point, there is \emph{no} way to separate minihalos from the IGM without resolving them because both appear in emission.  Interestingly, the treatment of the entropy floor makes essentially no difference to the results, because feedback kicks in only when the IGM is already strongly coupled (and usually saturated in emission).  

Thus minihalos are unlikely to have \emph{any} observable effect on the 21 cm signal (except on small spatial scales, obviously).  Of course, one can quibble with the particular parameters of our model.  But we have endeavored to choose values that maximize the importance of minihalos, by for example delaying reionization to $z \sim 7$ and ignoring other sources of fluctuations in $\dTb$.  Only by moving Ly$\alpha$ coupling closer to reionization would matters be improved.  Perhaps the best possibility is if very massive Population III stars dominate the radiation background throughout reionization.  These have much harder spectra than normal stars and so induce later \lya coupling \citep{furl06-glob}.  But we find that, even in this case, minihalos make only a small difference to the signal, and then only at $z \ga 12$.   Moreover, it seems implausible that star formation in pristine gas can dominate the ionizing photon budget \emph{throughout} reionization (see, e.g., \citealt{furl05-double}).  

\section{Discussion}
\label{disc}

We have presented a new method for calculating the 21 cm emission fluctuations from minihalos (via the halo model).  This made it particularly easy to estimate the effects of IGM heating (or an `entropy floor;' OH03) and allowed us to compare the minihalo and IGM fluctuation amplitudes in a self-consistent way.  Previous calculations that only included gravitational physics (and hence implicitly assumed both radiative heating and Ly$\alpha$ pumping of the IGM to be negligible) found that the 21 cm signal from minihalos dominates over that of the IGM \citep{shapiro05}.  In contrast, we found that the minihalo signal can only be unambiguously distinguished from the IGM in exceptional circumstances.  First, the Jeans mass must remain sufficiently low that small mass minihalos are able to form; once $T_K \ga 200 \kel$, the IGM signal becomes comparable to that of minihalos at $z=10$.  The required temperature is even smaller at higher redshifts.  Second, Wouthuysen-Field coupling must remain negligible.  Once the \lya background becomes significant, the diffuse IGM inevitably shines in either absorption or emission; because it contains so much more mass than the minihalo phase, it dominates the total signal \citep{oh03}.  

More importantly, even if the IGM does remain cold to relatively low redshifts $z\le 10$ (which is only marginally consistent with the \emph{WMAP} measurement, because heating almost certainly precedes reionization), {\it it is almost impossible to distinguish the minihalo contribution from that of an IGM with a slightly different reionization history}. Only if minihalos can be resolved (requiring arcsecond resolution) will they be robustly separable.\footnote{Nonlinear bias could ease the resolution requirements somewhat, if that can be modeled sufficiently precisely \citep{iliev03}.} Otherwise, the best way to infer the minihalo contribution is to measure both the absolute $\bdTb(z)$ and the large-scale fluctuation amplitude during a phase in which $T_K < T_\gamma$ (see Fig.~\ref{fig:pop2_noheat}).  Minihalos affect the mean signal only through their mass fraction, but their fluctuations are amplified by their bias.  This bias cannot be mimicked by a different ionization or thermal history and therefore presents a unique signature.  However, the cold phase almost certainly ends at $z \sim 15$ when $\fmh$ is still extremely small, so taking advantage of this separation my well be impossible in practice.  Once $T_S \ga T_\gamma$, both phases acquire the same emission characteristics and become indistinguishable.  In most histories, this later phase is by far the longest (and of course occurs preferentially at lower redshifts, when minihalos are more common).

Even if minihalos can be detected in this way (or if the diffuse IGM somehow maintains $T_S=T_\gamma$ for an extended period of time), we have shown that minihalos cannot be used to constrain cosmological parameters \citep{iliev02, iliev03} because of the inevitable astrophysical uncertainties in their emission properties.  By far the most important is the uncertain thermal history (and hence Jeans mass), which strongly affects $\fmh$ through the entropy floor.  Instead, in the unlikely event that they are seen, minihalos will constrain the astrophysical properties of the first luminous sources.

It is still possible to detect minihalos in other ways, and we certainly do not wish to downplay their significance for reionization \citep{iliev05, ciardi05-mh}.  The most promising possibility is the ``21 cm forest," in which minihalos appear as weak 21 cm absorption features toward a bright background radio source \citep{carilli02, furl02-21cm, furl06-forest}.  Because it is relatively easy to achieve high spectral resolution, individual minihalos can be observed without any degeneracies.  One particularly interesting application is to search for neutral regions (identified through their 21 cm emission, for example) that lack minihalo absorption lines.  These would presumably be hot regions of the IGM and may be relic \ion{H}{2} regions (OH03).  The main problem with this technique is identifying sufficiently bright background sources; models predict that they should exist to $z \ga 10$ but will become increasingly rare \citep{carilli02, haiman04}.  Another possibility is to search for non-gaussianities induced by the nonlinear gravitational clustering of minihalos, although this is likely to be difficult.  

\acknowledgements

We thank M. Zaldarriaga for helpful comments on the manuscript. We are also grateful to the Tapir group at Caltech for their hospitality while much of this work was completed. SPO acknowledges NSF grant AST-0407084 for support. 


\end{document}